# Data-Driven Personalized Energy Consumption Range Estimation for Plug-in Hybrid Electric Vehicles in Urban Traffic

Mehmet Fatih Ozkan*, James Farrell**, Marcello Telloni*, Luis Mendez**, Radu Pirvan*, Jeffrey P. Chrstos*, Marcello Canova**, Stephanie Stockar**

*Center for Automotive Research, The Ohio State University, Columbus, OH 43212 USA (e-mail: ozkan.25@osu.edu)

**Department of Mechanical and Aerospace Engineering, The Ohio State University, Columbus, OH 43210 USA

**Abstract**: In urban traffic environments, driver behaviors exhibit considerable diversity in vehicle operation, encompassing a range of acceleration and braking maneuvers as well as adherence to traffic regulations, such as speed limits. It is well-established that these intrinsic driving behaviors significantly influence vehicle energy consumption. Therefore, establishing a quantitative relationship between driver behavior and energy usage is essential for identifying energy-efficient driving practices and optimizing routes within urban traffic. This study introduces a data-driven approach to predict the equivalent fuel consumption of a plug-in hybrid electric vehicle (PHEV) based on an integrated model of driver behavior and vehicle energy consumption. Unlike traditional models that provide point predictions of fuel consumption, this approach uses Conformalized Quantile Regression (CQR) to offer prediction intervals that capture the variability and uncertainty in fuel consumption. These intervals reflect changes in fuel consumption, as well as variations in driver behavior, and vehicle and route conditions. To develop this model, driver-specific data were collected through a driver-in-the-loop simulator, which tested different human drivers' responses. The CQR model was then trained and validated using the experimental data from the driver-in-the-loop simulator, augmented by the synthetic data generated from Monte Carlo simulations conducted using a calibrated microscopic driver behavior and vehicle energy model. The CQR model was evaluated by comparing its predictions of equivalent fuel consumption intervals with those of baseline prediction interval methods that rely solely on conformal prediction.

## 1. INTRODUCTION

The rapidly increasing demand for personal mobility and the increase in vehicle miles traveled (VMT) have led to a substantial increase in energy consumption and greenhouse gas (GHG) emissions in recent years (Ma and Wang, 2022). This critical challenge motivates research and development efforts to promote energy conservation and sustainability in transportation (Kopelias et al., 2020).

Several strategies have been developed to mitigate the impact of the transportation sector on $CO_2$ emissions, including improvements in vehicle electrification technologies, connectivity, advanced driver assistance systems, and enhancements in traffic management systems. Among these, the analysis of human driver behavior's impact on vehicle energy use and emissions is an area that remains relatively underexplored. This is partly due to the complex nature of quantifying real-time driving behaviors and translating them into actionable data. Research shows that specific behaviors such as patterns of acceleration, consistent braking, and smooth deceleration can significantly influence the energy cost associated with different drivers. Analysis of driving data can be crucial for understanding the relationship between driving behaviors and vehicle energy consumption. This enables researchers to better quantify the direct effects of driver behavior on energy efficiency and propose targeted strategies to curb emissions, taking into account the human element in vehicular operations (Ping et al., 2019). Studies including (Ma and Wang, 2022) investigate the human driver in relation to vehicle fuel consumption by parameterizing driving characteristics as inputs to a Gaussian Process Regression (GPR) model and predicting vehicle fuel consumption to recommend fuel-friendly routes. While the model delivers accurate predictions of fuel consumption for an individual driver, inherent prediction variance is present. This uncertainty in fuel consumption is influenced by variations in driver behavior, as well as by vehicle and route characteristics. Such factors complicate point predictions for data-driven approaches. Consequently, a model that provides a prediction interval is necessary to more effectively represent the uncertainty in the data. Conformal prediction methods have been recently applied in the field of regression problems to address the point prediction limitation (Shafer and Vovk, 2008) (Manokhin, 2023). Conformalized quantile regression (CQR) is one of the conformal prediction models that provides an adaptable and resilient technique for uncertainty estimation (Romano et al., 2019). By leveraging the benefits of both conformal prediction and quantile regression frameworks, CQR can guarantee proper coverage probabilities, making it ideal for modeling scenarios where the variability of the response variable differs across different predictor values (Romano et al., 2019). CQR has been demonstrated successful for uncertainty estimation in a variety of domains including wind power forecasting (Jonkers et al., 2024), and vehicle trajectory prediction (Tumu et al., 2023).

This paper introduces a data-driven framework for predicting the equivalent fuel consumption range of human drivers using the CQR model, tailored to individual driving strategies and varying route conditions in urban traffic scenarios. The contribution of this work lies in establishing statistically valid prediction intervals that encapsulate the equivalent fuel consumption ranges, reflecting the unique behaviors of drivers and the dynamic nature of vehicle and route conditions.

## 2. DRIVER BEHAVIOR AND PHYSICAL VEHICLE ENERGY CONSUMPTION MODEL

The energy consumption prediction model utilizes the data of a Plug-in Hybrid Electric Vehicle (PHEV) integrated with a driver model. The system is composed of two main components: a deterministic driver model that simulates human driving behaviors and an energy consumption model that incorporates the vehicle energy management strategy, the vehicle longitudinal dynamics and the powertrain of the PHEV. The structure of the integrated model, highlighting the relationship between the human driver model and the PHEV model is shown in Fig. 1.

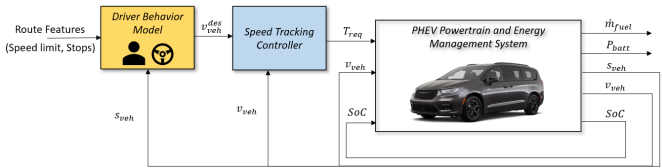

Fig. 1 Schematic of the integrated driver behavior and PHEV energy consumption model.

### 2.1 Driver Behavior Model

The behavior of a human driver is modeled considering the enhanced driver model (EDM) (Gupta et al., 2019). The EDM is a deterministic model that predicts vehicle speed by considering route features (such as speed limits and stop signs) and the presence of a lead vehicle. The EDM comprises three operational modes: Car Following (CF), Freeway Driving (FD), and Stop Mode (SM). The Freeway Driving mode becomes active when the driver's line of sight (LoS) extends up to 100 meters without encountering a lead vehicle or stop signs in preceding traffic. The Stop Mode activates when the distance to the stop sign decreases below the critical braking distance $x_{brake}$, guiding the vehicle to decelerate until it reaches a standstill at a safe gap $x_{safe}$ before the stop sign. The Car Following mode takes priority when both Freeway Driving and Stop Mode are enabled while a lead vehicle is detected ahead of the ego vehicle. The following equations detail the EDM and its different operating modes:

$$\frac{dv_{veh}(t)}{dt} = \begin{cases} a\left[1 - \left(\frac{v_{veh}(t)}{v_{\lim} - \theta}\right)^\delta\right], & \text{if } v_{veh}(t) \leq v_{\lim} \quad [FD] \\ -b\left[1 - \left(\frac{v_{\lim} - \theta}{v_{veh}(t)}\right)^\delta\right], & \text{if } v_{veh}(t) > v_{\lim} \quad [FD] \\ a\left[1 - \left(\frac{v_{veh}(t)}{v_{lead}(t)}\right)^\delta\right], & \text{if } v_{veh}(t) \leq v_{lead}(t) \quad [CF] \\ -b\left[1 - \left(\frac{v_{lead}(t)}{v_{veh}(t)}\right)^\delta\right], & \text{if } v_{veh}(t) > v_{lead}(t) \quad [CF] \\ -b^{-1}\left(\frac{v_{veh}(t)^2}{2s(t)}\right)^2, & \text{if } s(t) < s_{brake}(t) \quad [SM] \end{cases}$$
(1)

$$s(t) = x_{next}(t) - x_{ego}(t) - x_{safe} \quad (2)$$

$$s_{brake}(t) = \left(1 + \frac{c_1}{\delta}\right)\frac{v_{veh}(t)^2}{2b} \quad (3)$$

where $v_{veh}$ is the ego vehicle velocity, $a$ is the maximum acceleration, $b$ is the maximum deceleration, $\theta$ is the speed limit offset, $\delta$ is the acceleration exponent that represents the driver's aggressiveness, $c_1$ is the critical braking calibration parameter, $v_0$ is the speed limit, $x_{safe}$ is the safe following distance, $x_{ego}$ is the longitudinal position of the ego vehicle, $x_{next}$ is the distance of the ego vehicle toward the next obstacle in traffic (upcoming stop sign or lead vehicle). The key model parameters reflecting individual driving preferences are $[a, b, \delta, c_1, \theta]$. These parameters can be calibrated using experimental vehicle speed data to mimic the driving behaviors of human drivers, as reported by (Gupta et al., 2020).

### 2.2 Physics-Based Vehicle Energy Consumption Model

The EDM's ability to generate driver-specific speed trajectories enables direct coupling with a forward-looking vehicle energy simulator that predicts the behavior of the powertrain and the energy management system. This integration allows to examine how human drivers' individual driving responses along with the route conditions influence vehicle energy utilization for a given itinerary.

A forward-looking simulator of Chrysler Pacifica PHEV, previously calibrated to experimental data (Ozkan et al., 2024), was combined with the EDM, according to the block diagram shown in Fig. 1, where the EDM generates a reference velocity trajectory, which is followed by a gain-scheduled PI speed tracking controller. The controller generates the requested torque $T_{req}$ that is an input to the PHEV powertrain and energy management system. The energy consumption-related outputs of the PHEV powertrain and energy management block can be analyzed in terms of instantaneous fuel consumption $\dot{m}_{fuel}$, battery power $P_{batt}$ and state of charge $SoC$. The total equivalent fuel consumption of a given trip is computed as:

$$m_{f,eq} = m_{fuel} + \frac{1}{LHV}E_{batt} \quad (4)$$

where $m_{fuel}$ is the total fuel consumed, $E_{batt}$ is the total battery energy and the lower heating value $LHV = 42761\ J/g$

is obtained assuming the equivalence of 1 gallon of fuel to 33.7 kWh of battery electric energy (EPA, 2011).

## 3. DATA-DRIVEN VEHICLE ENERGY CONSUMPTION PREDICTION MODEL

The CQR framework is adopted to estimate the equivalent fuel consumption range of a PHEV by incorporating the driver behavior, vehicle and route conditions. Given training data $I_T = \{(X_1, Y_1), ..., (X_n, Y_n)\}$ which consists of $n$ pairs of the input $(X)$ and output $(Y)$ vectors, the primary objective of CQR is to construct prediction intervals that ensure a specified coverage probability for a new input vector $X_{n+1}$. This probability represents the likelihood that the new actual observation $Y_{n+1}$ falls within these intervals, meeting or exceeding a defined coverage rate $1 - \alpha$, i.e., $P\left(Y_{n+1} \in C_{n,\alpha}(X_{n+1})\right) \geq 1 - \alpha$ where $C_{n,\alpha}$ is the constructed prediction interval and $\alpha$ is the desired miscoverage rate. To construct the prediction interval $C_{n,\alpha}$, the CQR model requires a quantile regressor. The Light Gradient Boosting Machine (LightGBM) regressor (Ke, et al. 2017) was chosen as the quantile regressor for this purpose. LightGBM is a tree-based quantile regression algorithm, which takes advantage of the gradient boosting structure combined with a quantile loss function to estimate conditional quantiles of a response variable. This feature makes LightGBM ideal for capturing complicated dependencies and non-linear interactions in data, allowing for precise estimates of the appropriate conditional quantiles (Ke, et al. 2017). Moreover, LightGBM has the advantages of faster training, lower memory consumption, better accuracy and compatibility with large datasets compared to the existing gradient-boosting algorithms (Cao et al., 2023). In the LightGBM regressor, eight input features were considered to predict the equivalent fuel consumption $m_{f,eq}$ of a given itinerary. The input states were specifically selected to capture the human driver behavior, vehicle and route characteristics. Therefore, the input features of the LightGBM were defined as $X = [a, b, \delta, c_1, \theta, v_{lim}, l_r, SoC_0]$ where $a, b, \delta, c_1, \theta$ are the EDM parameters, $v_{lim}$ and $l_r$ are speed limit and length of the route itinerary, respectively, and $SoC_0$ is the initial battery state of charge at the beginning of the route itinerary.

Constructing the prediction interval of the CQR model follows several steps. First, the dataset is split into the training $I_T$ and calibration data $I_C$. Then, the LightGBM regressor fits two conditional lower $q_{\alpha,lo}$ and upper $q_{\alpha,hi}$ quantile functions on the training set $I_T$, given a certain coverage rate $1 - \alpha$. After that, the conformity scores for each data sample in the calibration data set $I_C$ are being calculated as:

$$E_i = \max\{q_{\alpha,lo}(X_i) - Y_i, Y_i - q_{\alpha,hi}(X_i)\}, \ i = 1,2,...,k \quad (5)$$

where $k$ is the total number data samples in the calibration data set $I_C$. Finally, the prediction interval for $Y_{n+1}$ can be constructed as:

$$C_{n,\alpha}(X_{n+1}) = [q_{\alpha,lo}(X_{n+1}) - Q_\alpha, q_{\alpha,hi}(X_{n+1}) + Q_\alpha] \quad (6)$$

where $Q_\alpha \coloneqq (1 - \alpha)(1 + 1/|I_C|)$-th empirical quantile of $E_i: i \in I_C$ which conformalizes the prediction interval. If a given data $(X_i, Y_i), i = 1, ..., n + 1$ are exchangeable, $P\left(Y_{n+1} \in C_{n,\alpha}(X_{n+1})\right) \geq 1 - \alpha$ is theoretically guaranteed as justified by (Romano et al., 2019).

It is worth noting that realistic velocity profiles and driver responses are captured using a driver-in-the-loop simulator to ensure authentic driving conditions. However, the CQR model requires a substantial dataset to be effective. To address this, the EDM, validated to accurately represent human driver responses, is used to generate synthetic data. This artificial data supplements the training and validation sets, enhancing the robustness of the training process. Furthermore, this syntenic data is assumed to be exchangeable since the data was generated from diverse driving conditions, ensuring a common underlying distribution for both calibration and test datasets. Additionally, there is no temporal dependency among the data points, which aligns with the principles of conformal prediction where exchangeability is crucial for valid prediction intervals (Romano et al., 2019).

## 4. CASE STUDY

### 4.1 Driving Simulator Setup

An experimental study was conducted using the Vehicle Dynamics Driver-in-the-Loop (VDDiL) simulator at the Ohio State University Center for Automotive Research (OSU-CAR). The design, implementation, and validation of this simulator were previously undertaken as detailed in (Sekar et al., 2022). The VDDiL is an advanced driving simulator designed for vehicle dynamics testing and to offer a comprehensive platform for driver-in-the-loop experiments. The system features a D-BOX actuation system capable of delivering three degrees of freedom motion (roll, pitch, and heave), coupled with a SENSO-Wheel SensoDrive for realistic steering torque feedback. Pedals provide control over acceleration and braking, with the brake pedal endowed with an anti-lock braking system (ABS) that generates pulses when engaged. Immersive realism was achieved through a combination of a real vehicle cockpit, a three-display setup, and an audio system. Furthermore, SCANeR Studio by AV Simulation facilitates the creation of virtual environments and controlled traffic simulations, enhancing the system's adaptability and suitability for various vehicle development and testing purposes. The reader is referred to (Sekar et al., 2022) for further details on the VDDiL simulator development.

### 4.2 Driver Data Collection

Driver-specific data were collected using an 8 km simulated urban route selected from Columbus, OH, (Fig. 2) to emulate real-world urban conditions typical of daily commuting. The route was divided into two segments, each representing different driving scenarios. The first segment is an urban environment, with turns, traffic lights, stops, and varied speed limits. The second segment corresponds to a highway scenario, characterized by higher speeds. The fidelity of the reconstructed road features, closely resembling those of a real-world route, enhances the experimental validity and instills confidence in drivers regarding the surrounding scenario and road conditions (Sekar et al., 2022). Data collection involved 26 participants who drove through the route aiming to replicate their real-world driving behavior as accurately as possible.

Participants were instructed to follow speed limits and drive as they normally would. To ensure natural driving behavior and adherence to traffic regulations, drivers received real-time feedback on their speed and the road speed limits during the test, as shown in Fig. 3. All traffic lights on the route were kept green to avoid inconsistencies due to traffic light timings.

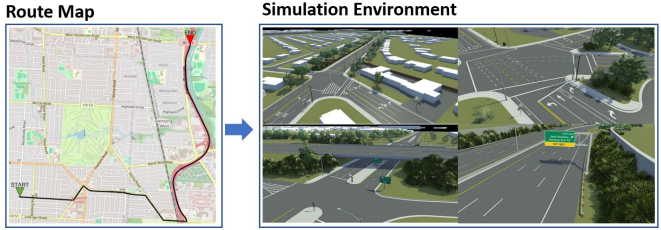

Fig. 2 Urban route map and simulation environment.

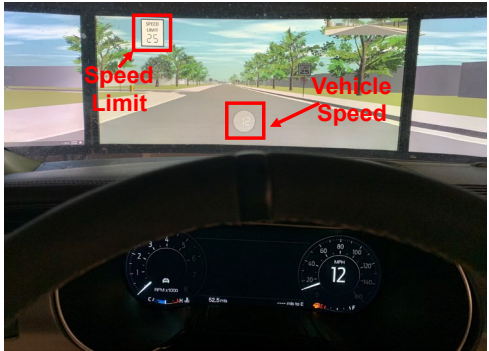

Fig. 3 A driving scene from the driver's view.

*4.3 Driver Behavior Model Development and Data Generation*

The calibration of the EDM parameters was performed with the data collected from the VDDiL simulator. First, the route was partitioned into eight segments to identify the drivers' unique actions on different parts of the trip, such as approaching stop signs, adapting to speed limit changes, and driving on a multi-lane highway. Second, the EDM parameters were calibrated for each driver's test data across the respective itinerary using Genetic Algorithm (Gupta et al., 2020). A total of 208 calibrated EDM parameter combinations (8 parameter combinations for each driver) were achieved at the end of the calibration process. Fig. 4 illustrates the calibration results for one driver and route segmentation, along with the route characteristics. The calibrated EDM parameter, as depicted in Fig. 4, captures the overall behavior of the driver-specific velocity profile.

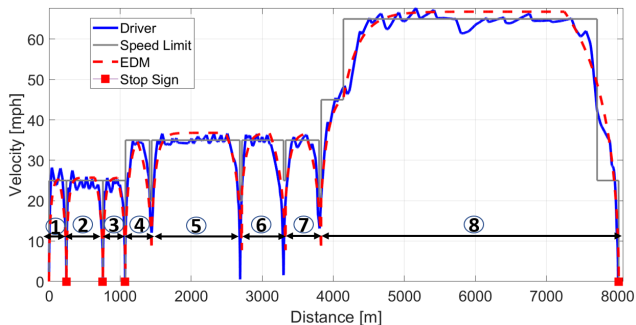

Fig. 4 Example of one driver's speed profile and route segmentation.

To further evaluate the performance of the EDM over the dataset, Root Mean Square Error (RMSE) is computed between the predicted speed and the actual speed trajectories, and the distribution of the prediction errors is shown in Fig. 5. The mean speed prediction error is 2.19 mph, while the majority of the prediction errors are less than 3 mph. The distribution of speed prediction errors agrees with that reported by (Gupta et al., 2022).

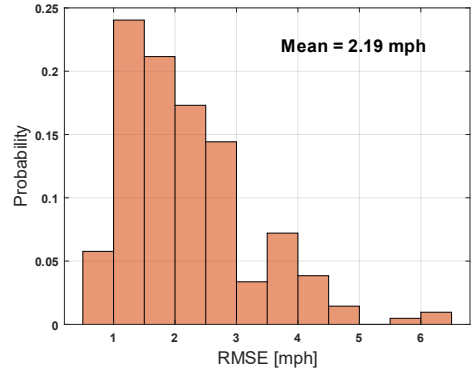

Fig. 5 EDM speed prediction error distribution.

The distribution of the calibrated EDM parameters across all collected datasets is shown in Fig. 6, revealing a distinct pattern of variability within the driving dataset. Notably, the parameters reflect specific behaviors observed among drivers. For instance, a considerable number of drivers exceeded the speed limits (indicated by negative $\theta$ values), while most demonstrated a tendency for aggressive acceleration and more relaxed deceleration, as shown by the distribution of maximum acceleration ($a$) and deceleration ($b$) parameters. These observations collectively showcase the diverse range of driving styles captured in the dataset.

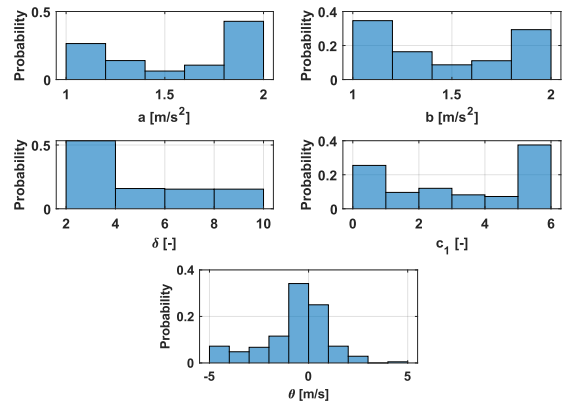

Fig. 6 Distribution of the calibrated EDM parameters.

A Monte Carlo simulation was then performed to augment the experimental data set for creating the CQR-based equivalent fuel consumption estimator. A t-copula multivariate distribution model (Bouye et al., 2000) was fitted to the calibrated EDM parameters to generate a distribution of parameter values reflecting the true underlying variability and dependency structure (Ozkan et al., 2021) (Ozkan and Ma, 2022). Next, 1000 sets of EDM parameters were randomly drawn from the fitted t-copula distribution to generate the synthetic data necessary for training. For comparison, Fig. 7 overlays the distribution of EDM parameters from test data with synthetic values generated by the t-copula model. This demonstrates that synthetic parameters effectively capture the overall distribution of the original EDM parameters. These copula-generated parameters were then used to simulate driver-specific trajectories across eight route segments, starting with initial $SoC$ values of 26%, 30%, and 40%.

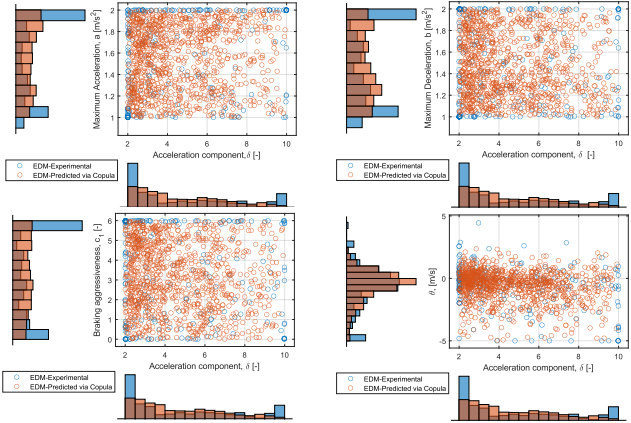

Fig. 7 Distribution of the EDM parameter obtained from drivers' data and EDM parameters generated from the t-copula distribution.

Table 1 The tuned hyperparameters of the LightGBM regressor.

| Parameter | Value |
|---|---|
| Number of boosted trees | 955 |
| Learning rate | 0.19 |
| Maximum tree depth for base learners | 14 |
| Maximum tree leaves for base learners | 42 |

## 5. RESULTS AND DISCUSSION

### 5.1 Training Methodology

The CQR model was created from a total of 24000 data points, obtained from the generated trajectories from the Monte Carlo simulation. The training, validation, and testing data sizes for the CQR model are set to 80%, 10%, and 10% of the data, respectively and the desired miscoverage rate $\alpha$ is set to 0.1. To assess the effectiveness of the CQR approach, the prediction performance of the CQR model was analyzed with baseline state-of-the-art conformal prediction methods including CV and CV+ (Barber, et al., 2021), and Jackknife+-after-Bootstrap (Kim, et al., 2020). The baseline methods are developed with the same data set and split used in CQR development and the desired miscoverage rate for the baseline methods is also set to 0.1. In the CQR and the baseline methods, the hyperparameters of the LightGBM quantile regressor were tuned by using the RandomizedSearchCV function in (Pedregosa et al, 2011) to optimize the model in order to reduce the prediction error. The calibrated hyperparameters of the LightGBM were listed in Table 1. To train and test the CQR and baseline methods, the Model Agnostic Prediction Interval (MAPIE) open-source Python library (Taquet et al., 2022) was used.

### 5.2 Evaluation of the CQR Model and Baseline Methods

Fig. 8 compares predicted and simulated equivalent fuel consumption results using CQR and baseline methods. Note that only 10% of observations are shown for clarity. Both methods achieve a coverage rate above 90%, with CQR demonstrating narrower prediction intervals on average compared to all baseline methods.

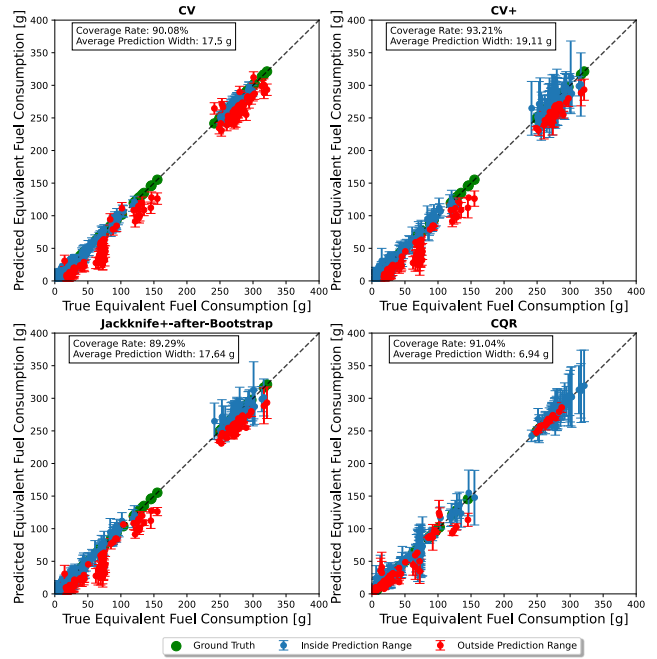

Fig. 8 Equivalent fuel consumption range estimation results with the proposed CQR and baseline CV, CV+ and Jackknife+-after-Bootstratp methods.

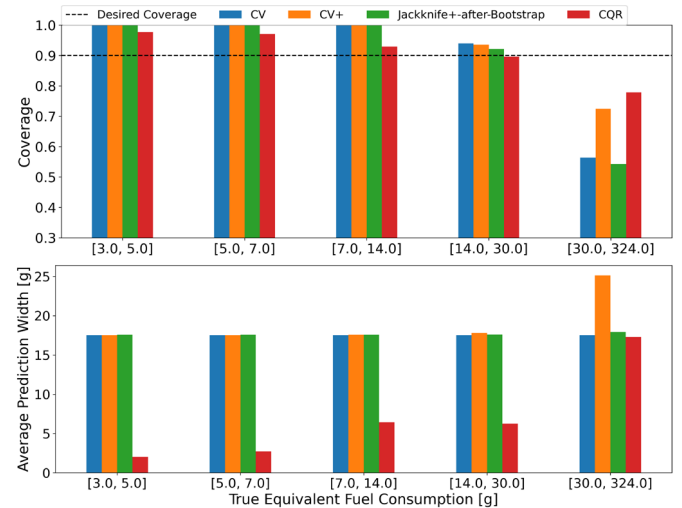

Fig. 9 Coverage and average prediction range comparison of the proposed CQR and baseline CV, CV+ and Jackknife+-after-Bootstratp methods.

To further investigate the performance of the CQR and baseline methods, Fig. 9 shows the resulting coverage and prediction range comparison on the prediction intervals split by quintiles. Greater adaptability in the prediction intervals was observed with the CQR method compared to baseline methods that employed a fixed prediction range. For instance, CQR provides narrower prediction widths for the regions where the true equivalent fuel consumption is low, and the prediction widths computed by CQR expand as true equivalent fuel consumption values increase. The primary reason behind this observation is that CQR can adaptively provide narrower and wider prediction widths in regions of the data distribution where uncertainty is lower and higher, respectively. This adaptability enhances prediction precision by aligning widths with varying levels of uncertainty in the data. The statistical results demonstrate that the CQR method's adaptability and precision make it a promising approach for predicting PHEV equivalent fuel consumption based on a human's unique driving style. It offers improved coverage rates and narrower prediction intervals compared to state-of-the-art methods. These benefits can enhance route planning, fuel management, and budgeting for both individual drivers and commercial fleets, optimizing fuel efficiency and cost accuracy.

## 6. CONCLUSIONS AND FUTURE WORK

In this paper, a data-driven equivalent fuel consumption prediction model for PHEV was developed using the CQR approach. A human driver behavior model was calibrated using data from a comprehensive human subject study with a driver-in-the-loop simulator. This calibrated behavior model, combined with a physics-based PHEV energy consumption model, formed the basis for training the CQR model. The CQR approach predicts equivalent fuel consumption on urban routes, considering individual driving styles, vehicle characteristics, and route conditions. Results show that CQR provides accurate fuel consumption intervals, outperforming baseline conformal prediction methods by combining quantile regression and conformal prediction. Future work will explore online driver parameter estimation methods and personalized eco-driving strategies for advanced driving assistance systems to enhance traffic energy efficiency.

## ACKNOWLEDGEMENTS

The authors are grateful to Stellantis for supporting this work and for the fruitful discussions that enhanced the outcomes of this research.